\documentclass[%
 reprint,
 amsmath,amssymb,
 aps,
 prd,
]{revtex4-2}

\usepackage{booktabs}  
\usepackage{makecell}
\usepackage{graphicx}
\usepackage{dcolumn}
\usepackage{bm}
\usepackage{wrapfig}
\usepackage{xcolor}
\usepackage{tabularx}
\usepackage{hyperref}

\begin{document}

\newcommand{\Mstar}{M_*}
\newcommand{\Mhole}{M_h}
\newcommand{\maya}{\textit{Maya}}
\newcommand{\etk}{\textit{Einstein Toolkit}}
\newcommand{\PL}[1]{{\textcolor{red}{\sf{[PL: #1]}}}}
\newcommand{\JD}[1]{{\textcolor{magenta}{\sf{[JD: #1]}}}}

\def\tov#1{Tolman–Oppenheimer–Volkoff#1 (TOV#1)\gdef\tov{TOV}}
\def\eos#1{equation of state#1 (EOS#1)\gdef\eos{EOS}}
\def\qnm#1{quasi-normal mode#1 (QNM#1)\gdef\qnm{QNM}}
\def\gw#1{gravitational wave#1 (GW#1)\gdef\gw{GW}}
\def\ns#1{neutron star#1 (NS#1)\gdef\ns{NS}}
\def\bs#1{boson star#1 (BS#1)\gdef\bs{BS}}
\def\bh#1{black hole#1 (BH#1)\gdef\bh{BH}}
\def\bbh#1{binary black hole#1 (BBH#1)\gdef\bbh{BBH}}
\def\bns#1{binary neutron star#1 (BNS#1)\gdef\bns{BNS}}
\def\bbs#1{binary boson star#1 (BBS#1)\gdef\bbs{BBS}}
\def\bhns#1{black hole - neutron star#1 (BHNS#1)\gdef\bns{BHNS}}
\def\bhbs#1{black hole - boson star#1 (BHBS#1)\gdef\bhbs{BHBS}}
\def\bsns#1{boson star - neutron star#1 (BSNS#1)\gdef\bsns{BSNS}}
\def\nr#1{numerical relativity#1 (NR#1)\gdef\nr{NR}}
\def\ligo#1{Laser Interferometer Gravitational-wave Observatory#1 (LIGO#1)\gdef\ligo{LIGO}}


\title{Head-on Collisions of Boson Stars with Bowen-York Type Initial Data}
\author{Jake Doherty and Pablo Laguna}
\affiliation{
Center for Gravitational Physics, Weinberg Institute, Department of Physics, The University of Texas at Austin, Austin, TX 78712, USA
}

\begin{abstract}
 We present a numerical relativity study of head-on collisions involving boson stars using initial data inspired by the Bowen-York initial data used to model black hole binaries with punctures. The initial data method preserves the simplicity of the Bowen-York approach, thus allowing incorporating information from the early, post-Newtonian inspiral phase in binary coalescences. We test the method on a single boson star with linear momentum. We present results from head-on collisions of boson stars as well as encounters of boson stars with black holes. In general, the results are consistent with previous studies, demonstrating the effectiveness of the initial data method. In particular, we show that boson star head-on collisions emit more energy in gravitational waves than the equivalent black hole binaries. On the other hand, head-on collisions of a boson star with a black hole radiate less than their black hole binary counterparts.  
\end{abstract}

\maketitle

\section{Introduction}

There is a rich literature on \bs{s} that spans back to the 1960s. Building off Wheeler's attempt to construct general relativistic self-gravitating objects from the electromagnetic field \cite{Wheeler1955}, Kaup proceeded to do so via a massive complex scalar field \cite{Kaup1968}, giving birth to the concept of \bs{s}. Since then there has been a wide variety of general relativistic numerical studies into their behavior, such as (to name a few) stability analyses of the stationary star \cite{Seidel1990, Kain2021}, critical phenomena of single stars \cite{HawleyChoptuik2000, Rousseau2003, Lai2004, Esco2017, Alcubierre2019}, head-on collisions \cite{Palenzuela:2006wp, Choptuik2010, mundim2010, Helfer2022, Evstafyeva2022, Jaramillo2022, Atteneder2024,Ge2025, Brito2026, Palloni2026}, and binary inspirals \cite{Palenzuela2007, Palenzuela2017, Bezares2017, Bezares2022, Siemonsen2023_0, Croft2023, Siemonsen2023_1, Evstafyeva2026} of two stars.

Compact-object mixed binaries involving \bs{s} have received much less attention in the literature. Dietrich et al. performed a numerical study of \bsns{} binaries, both head-on collisions and in-spiral mergers \cite{Dietrich_2019}. For \bhbs{} binaries, the piercing of a \bs{} by a small \bh{} was investigated in Ref.~\cite{Cardoso2022,Zhong2023}, and it was found that the \bh{} accretes almost the entire \bs{} even for the most extreme mass ratios. Very recently, head-on collisions and in-spirals of \bhbs{} systems were numerically simulated \cite{Marks2026}, and it was shown that the scalar field significantly affects the radiative efficiency of the system and that \bhbs{} binaries always radiate less energy than their equivalent \bbh{} counterparts. Very soon after this, another study into head-on collisions of \bhbs{} appeared~\cite{Ning2026}, and found that higher multipoles in the gravitational radiation can help distinguish \bhbs{} binaries from their \bbh{} counterparts. 

Currently, the most common approach to constructing initial data with \bs{s} is to superpose boosted solutions for isolated stars \cite{Helfer2022}. This initial data is only an approximate solution since it does not satisfy the constraints. However, the evolution scheme is such that the violations decay away~\cite{Palenzuela2017}. Other approaches solved the constraint equations using a conformal thin-sandwich decomposition. This was done in Ref.~\cite{Dietrich_2019} for the case of \bsns{} binaries and in \cite{Siemonsen2023_1} for generic \bbs{} binaries. Ref.~\cite{Atteneder2024} provides a detailed review and numerical analysis of these initial data approaches and their subsequent evolutions. Finally, and also recently, Ref.~\cite{Palloni2026} constructed constraint-satisfying initial data for \bbs{} binaries through the so-called extended conformally flat formalism. 

The present study focuses on a new approach to construct initial data for numerical relativity simulations involving \bs{s}. The study considers only head-on collisions between two \bs{s} or between a \bs{} and a \bh{}. The new initial data method is inspired by the Bowen-York approach to \bbh{}. It preserves its simplicity, which allows for a natural connection to post-Newtonian approximations during the early inspiral phase of a binary coalescence. We test the method on a single \bs{} with linear momentum.  In general, our head-on collision results are consistent with previous studies. In particular, \bbs{} systems emit more energy in \gw{s} than the equivalent \bbh{} systems. On the other hand, \bhbs{} head-on collisions always radiate less energy than their binary \bh{} counterparts. 

The paper is organized as follows: A summary of the conformal transverse traceless approach to initial data is provided in Section~\ref{sec:init_data}. Solutions to the momentum constraint for both perfect fluid sources and \bs{s} are derived in Section~\ref{sec:solve_mom_const}. Tests of a single \bs{} with linear momentum are given in Section~\ref{sec:SingleBS}. Results of equal mass \bbs{} head-on collisions are shown in Section~\ref{sec:BBS}.  Section~\ref{sec:BHBS} presents results from \bhbs{} head-on collisions and comparisons with \bbh{} and \bbs{} encounters. Conclusions are given in  Section~\ref{sec:conclusions}. We use units in which $G=c=1$. Latin indices from the beginning of the alphabet denote space-time indices, and those from the middle denote spatial indices. All numerical results are reported in units of $\mu$, the mass of the complex scalar field. All simulations were performed with the \maya{} code \cite{2003VPPR5ICGoodale, Husa2006, 2012ApJHaas, 2015ApJLEvans,2016PRDClark,2016CQGJani}, our local version of the \etk{} code \cite{maxwell_rizzo_2025_15520463}. The code is based on the BSSN formulation~\cite{baumgarte_shapiro_2010} and uses the moving puncture gauge conditions.

\section{Conformal-Transverse-Traceless Approach to Initial Data}
\label{sec:init_data}
When viewed as an initial value problem, the Einstein equations of General Relativity are expressed in terms of a 3+1 foliation of constant time $t$, space-like hypersurfaces $\Sigma_t$ with unit normal $n^a$. Under this splitting, the space-time line element takes the form:
\begin{eqnarray}
    ds^2 &=& -\alpha^2\,dt^2 + \gamma_{ij}(dx^i+\beta^i)(dx^j+ \beta^j) \,,\label{eq:metric}
\end{eqnarray}
where $\gamma_{ij}$ is the intrinsic metric to $\Sigma_t$, and $\alpha$ and $\beta^i$, the lapse and shift connecting slices in the foliation. In terms of $n^a$ and the space-time metric $g_{ab}$, the intrinsic metric is also given by $\gamma_{ab} = g_{ab}+n_a\,n_b$, where $n^a = (1,-\beta^i)/\alpha$. Using the spatial metric $\gamma_{ab}$ and the normal $n^a$, the Einstein equations can be projected into $\Sigma_t$ and along $n^a$. Projecting the equations twice along $n^a$ and one long $n^a$ and one into  $\Sigma_t$, one obtains the constraint equations that the initial data must satisfy. Specifically,
\begin{eqnarray}
\label{eq:ham}
    R + K^2 - K_{ij} K^{ij} &= 16 \pi \rho_H\,\\
\label{eq:mom}
    \nabla_j \left(K^{ij} - \gamma^{ij}K \right) &= 8 \pi S^i,
\end{eqnarray}
which are, respectively, the Hamiltonian and momentum constraints. In these equations, $R$ is the Ricci scalar, $\nabla_j$ is the covariant derivative associated with $\gamma_{ij}$, and $K_{ij}$ is the extrinsic curvature, where $K = K^i\,_i$. 
The matter sources are:
\begin{eqnarray}
\label{eq:hmdensity_pf}
\rho_H &\equiv& n^a n^b T_{ab}\\
\label{eq:momdensity_pf}
S^i &\equiv& -\gamma^{ij} n^b T_{jb}\,,
\end{eqnarray}
where $T_{ab}$ is the stress-energy tensor, $\rho_H$ is the mass-energy density, and $S^i$ is the momentum density. 

A popular approach to identify the elements in the initial data $\lbrace \gamma_{ij}, K_{ij}\rbrace$ that the constraints fix is using the conformal-transverse-traceless (CTT) decomposition pioneered by Lichnerowicz, York, and collaborators~\cite{baumgarte_shapiro_2010,Smarr:1979ofa} in which 
\begin{eqnarray}
\label{eq:confgam}
    \gamma_{ij} &=& \psi^4 \tilde{\gamma}_{ij}\\
\label{eq:confKij}
    K_{ij} &=& A_{ij} + \frac{1}{3}\gamma_{ij}K \\
\label{eq:confAij}
    A^{ij} &=&\psi^{-10} \tilde{A}^{ij}\\
\label{eq:confSi}
     S^i &=&  \psi^{-10}\tilde S^i\,.
\end{eqnarray}
Also a common practice is to impose conformal flatness ($\tilde{\gamma}_{ij} = \eta_{ij}$), and maximal slicing ($K=0$). Thus, the Hamiltonian and momentum constraints become:
\begin{eqnarray}
    \label{eq:ham2}
    &\tilde{\Delta} \psi+ \frac{1}{8}\psi^{-7}\tilde{A}_{ij}\tilde{A}^{ij} = -2\pi \psi^{5}\rho_H \\
    \label{eq:mom2}
    &\widetilde{\nabla}_j \tilde A^{ij} = 8 \pi \tilde{S}^i.
\end{eqnarray}
Notice that with these assumptions,  the momentum constraint decouples from the Hamiltonian constraint. Thus, the first task is to find solutions to the momentum constraint in the form of Eq.~(\ref{eq:mom2}).
\section{Solving the Momentum Constraint}
\label{sec:solve_mom_const}
\subsection{Perfect Fluid Stars}

Bowen and York ~\cite{1980PhRvD..21.2047B} found point-source solutions to $\widetilde{\nabla}_j \tilde A^{ij} =0$ that have been extensively used to construct initial data with \bh{s} modeled as punctures~\cite{baumgarte_shapiro_2010}. The solution representing a \bh{} with linear momentum $P^i$ is given by:
\begin{eqnarray}
  \tilde A^{ij} &=& \frac{3}{2\,r^2}\left[ 2\,P^{(i} l^{j)} - ( \eta^{ij}-l^il^j)P_k l^k\right]\label{eq:KP}
\end{eqnarray}
where $l^i = x^i/r$ is a unit radial vector. Bowen~\cite{Bowen1979} found also a solution to Eq.~(\ref{eq:mom2}) for a spherically symmetric source $\widetilde S^i = P^i\,\sigma(r)$ representing an extended object with radius $R_*$ and linear momentum $P^i$. The solution is given by:
\begin{eqnarray}
\label{eq:Aij1}
\tilde{A}^{ij} &=& \frac{3Q}{2r^2}\left[2P^{(i}l^{j)} - (\eta^{ij} - l^il^j)P_kl^k  \right] \nonumber\\
&+& \frac{3C}{r^4}\left[2P^{(i}l^{j)} + (\eta^{ij} - 5l^il^j)P_kl^k  \right]\label{eq:KP2}
\end{eqnarray}
where
\begin{eqnarray}
	\label{eq:qjcdefs}
	Q &=&4\pi \int_0^r  \sigma \bar r^2 \, d\bar r \\
	C &=& \frac{2\pi}{3} \int_0^r  \sigma \bar r^4 \, d\bar r\,.
\end{eqnarray}
The source function $\sigma(r)$ has compact support in $r\le R_*$. Outside the source, $Q=1$ and $C=0$, thus the solution (\ref{eq:KP2}) reduces to the vacuum solution (\ref{eq:KP}). To ensure that $P^i$ is the momentum, the source function must satisfy the normalization condition
$\int\sigma\,dV = 1$.

In Ref.~\cite{Clark:2016ppe}, we applied the solution to a perfect fluid source.  In this case, the stress-energy tensor is given by
\begin{equation}
\label{eq:stressenergy}
T_{ab} = (\rho + p)u_a u_b + p\, g_{ab},
\end{equation}
with $\rho$ the total energy density, $p$ the pressure, and $u^a$ the four velocity of the fluid. With this, Eqs.~(\ref{eq:hmdensity_pf}) and (\ref{eq:momdensity_pf}) become
\begin{eqnarray}
\label{eq:hmdensity_pf2}
\rho_H &=&  (\rho + p)\,W^2 - p\\
\label{eq:momdensity_pf2}
S^i &=&  (\rho + p) \,W\,u^i,
\end{eqnarray}
where $W = -n_au^a$ is the Lorentz factor. A natural choice of conformal transformations that preserve the form of Eq.~(\ref{eq:confSi}) are $\tilde{\rho} = \rho\, \psi^8$, $\tilde{p} = p\, \psi^8$, $\tilde{u}^i = u^i\psi^2$ and $\widetilde W = W$. Thus, from Eq.~(\ref{eq:momdensity_pf2}), $\sigma\,P^i =  (\tilde\rho + \tilde p) \,W\,\tilde u^i$. Setting, $\sigma  = (\tilde\rho+\tilde p)/\mathcal K$ with ${\mathcal K}$ a constant, we get that $P^i = {\mathcal K}\,W\,\tilde u^i$ where
\begin{eqnarray}
 \mathcal K   &=&  4\,\pi \int_{0}^{R_*} (\tilde\rho+\tilde p)\,r^2\, dr\,.
\end{eqnarray}

\subsection{Boson Stars}

We are interested in using the Bowen solution (\ref{eq:KP2}) to the momentum constraint, but applied to \bs{s}. The stress-energy tensor for this case is
\begin{eqnarray}
    \label{eq:Tab_scalar}
     T_{ab} &=& \nabla_{(a}\Phi\, \nabla_{b)}\Phi^* - \frac{1}{2}g_{ab} \left( \nabla^c \Phi\, \nabla_c \Phi^* + 2\,V\right).
\end{eqnarray}
where $\Phi$ is a complex scalar field and $V$ its potential. A common choice is the \emph{solitonic} \bs{} potential
\begin{eqnarray}
    V &=& \frac{1}{2}\mu^2 |\Phi|^2\left(1-2\,\frac{|\Phi|^2}{\sigma_0^2}\right)^2  \,.
\end{eqnarray}
Here, we will concentrate on the \emph{mini} \bs{} potential. That is, when $\sigma_0 \rightarrow \infty$. With these choices, 
\begin{eqnarray}
\label{eq:Scalarrhoh}
    \rho_{H} &=& \frac{1}{2} \left( \Pi\, \Pi^* + \nabla^i \Phi \,\nabla_i \Phi^*\right) + \frac{1}{2} \mu^2 |\Phi|^2 \\
\label{eq:Scalarsi}
    S^i &=& \text{Re} \left[ \Pi^* \nabla^i \Phi\right]
\end{eqnarray}
where $\Pi = -n^a \nabla_a\Phi$. In order to preserve the conformal transformation $\tilde{S}^i = S^i \psi^{10}$, we impose $\Pi = \widetilde \Pi \,\psi^{-6}$ and $\Phi = \widetilde\Phi$. Thus,
\begin{eqnarray}
\label{eq:Scalarrhoh2}
    \rho_{H} &=& \frac{1}{2}\widetilde\Pi\, \widetilde\Pi^*\,\psi^{-12} + \frac{\psi^{-4}}{2}\widetilde\nabla^i \widetilde\Phi \,\widetilde\nabla_i \widetilde\Phi^* + \frac{1}{2} \mu^2 |\widetilde\Phi|^2\\
    \label{eq:Scalarsi2}
    \widetilde S^i &=& \text{Re} \left[ \widetilde\Pi^* \widetilde\nabla^i \widetilde\Phi\right]
\end{eqnarray}

We assume that initially the scalar field has the following structure:
\begin{eqnarray}
    \widetilde\Phi &=& \varphi(r)\,e^{i\left(\omega\,t - k^jx_j\right)}\,.\label{eq:anzats}
\end{eqnarray}
Therefore, 
\begin{eqnarray}
    \widetilde\Pi &=&  \frac{-1}{\tilde \alpha} \left(\partial_t - \beta^i \partial_i \right)\tilde \Phi =- \frac{i\,\omega}{\tilde\alpha} \widetilde\Phi\,,\label{eq:dot}
\end{eqnarray}
where we have set $\tilde\alpha=\psi^{-6}\alpha$ and $\beta^i=0$. In addition,
\begin{eqnarray}
    \widetilde\nabla^j\widetilde\Phi &=& \left( l^j\,\frac{\varphi'}{\varphi} - i\, k^j \right)\widetilde\Phi\,, \label{eq:grad}
\end{eqnarray}
where prime denotes derivatives with respect to $r$. Substitution of Eqs.~(\ref{eq:dot}) and (\ref{eq:grad}) into Eq.~(\ref{eq:Scalarsi2}) yields
\begin{eqnarray}
   \widetilde{S}^i &=& \frac{ \omega \, \varphi^2}{\tilde\alpha}k^i \,.
\end{eqnarray}
Since $\widetilde S^i = \sigma\,P^i$, we get that
\begin{eqnarray}
	P^{i} &=& {\mathcal{M}}\,k^i \\
    \sigma &=& \frac{\omega\,\varphi^2}{\mathcal{M}\,\tilde\alpha}\,,
\end{eqnarray}
where
\begin{eqnarray}
       \mathcal{M} &=& 4\,\pi\int_0^{\infty}\frac{\omega \,\varphi^2}{\tilde\alpha}\,r^2\,dr\,\label{eq:calM}.
\end{eqnarray}
With this in hand, the only input besides the momentum $P^i$ is the scalar field $\varphi(r)$ and the lapse function $\tilde\alpha(r)$. We obtain these functions from stationary \bs{} solutions in isotropic coordinates~ \cite{Ge2025}. 
 
\section{Single Boson Star with Linear Momentum}
\label{sec:SingleBS}

As a test, we first present results from evolving a single \bs{} with linear momentum.
For a given scalar field potential, spherically symmetric, stationary \bs{} solutions are uniquely characterized by the central amplitude $\varphi_*$ of the scalar field. This value also fixes the mass-energy $\Mstar$ of the star. Strictly speaking, the \bs{} stationary solution does not have a surface. The amplitude of the scalar field decays exponentially. A common convention is to define the surface of the \bs{} at the areal radius $R_*$ containing $99 \%$ of its mass. Mini-\bs{s} are stable for $\varphi_* \le 0.08$, so we set $\varphi_*=0.05$, which yields eigen-frequency $\omega=0.895\,\mu$, mass $\Mstar = 0.610\,\mu^{-1}$, and radius $R_*=10.22\,\mu^{-1}$. We endow the star with linear momentum along the $x$ axis within the range $0 \leq P/\Mstar \leq 0.3$.

\begin{figure}[!htbp]
\includegraphics[angle=0,width=0.4 \textwidth]{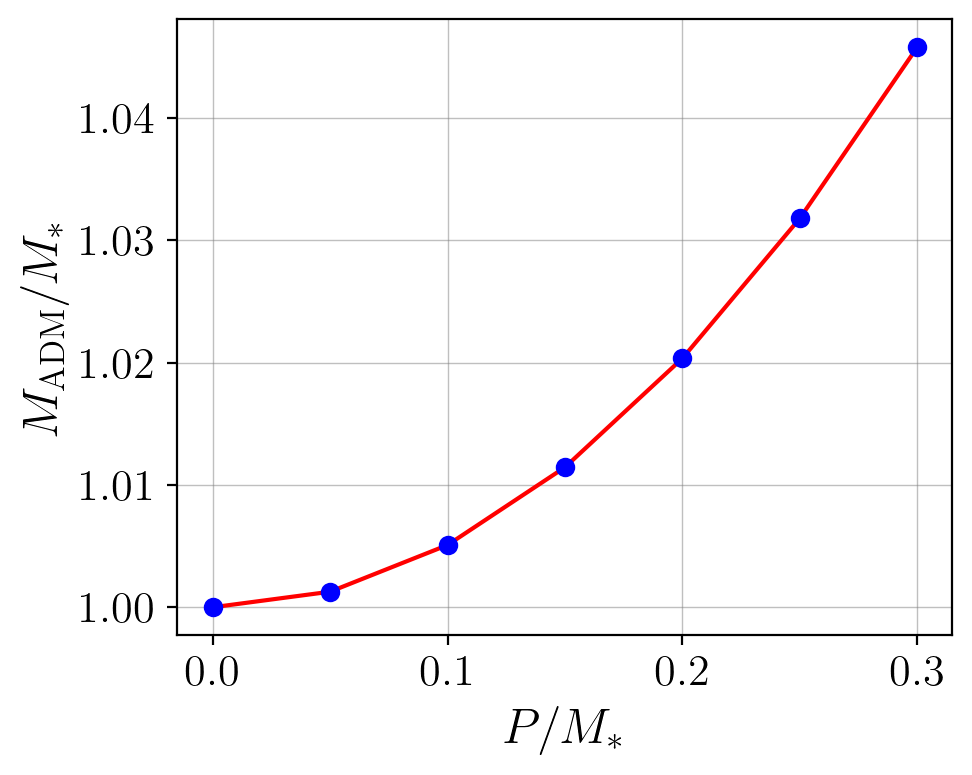}\\
\caption{ADM mass $M_{\text{ADM}}/\Mstar$ (dots) as a function of $P/\Mstar$ for a single \bs{}. The solid line represents a fit to $M_{\text{ADM}}/\Mstar = 1 + c(P/\Mstar)^2$.}
\label{fig:ADMmass}
\end{figure}

\begin{figure}[!htbp]
\includegraphics[angle=0,width=0.45 \textwidth]{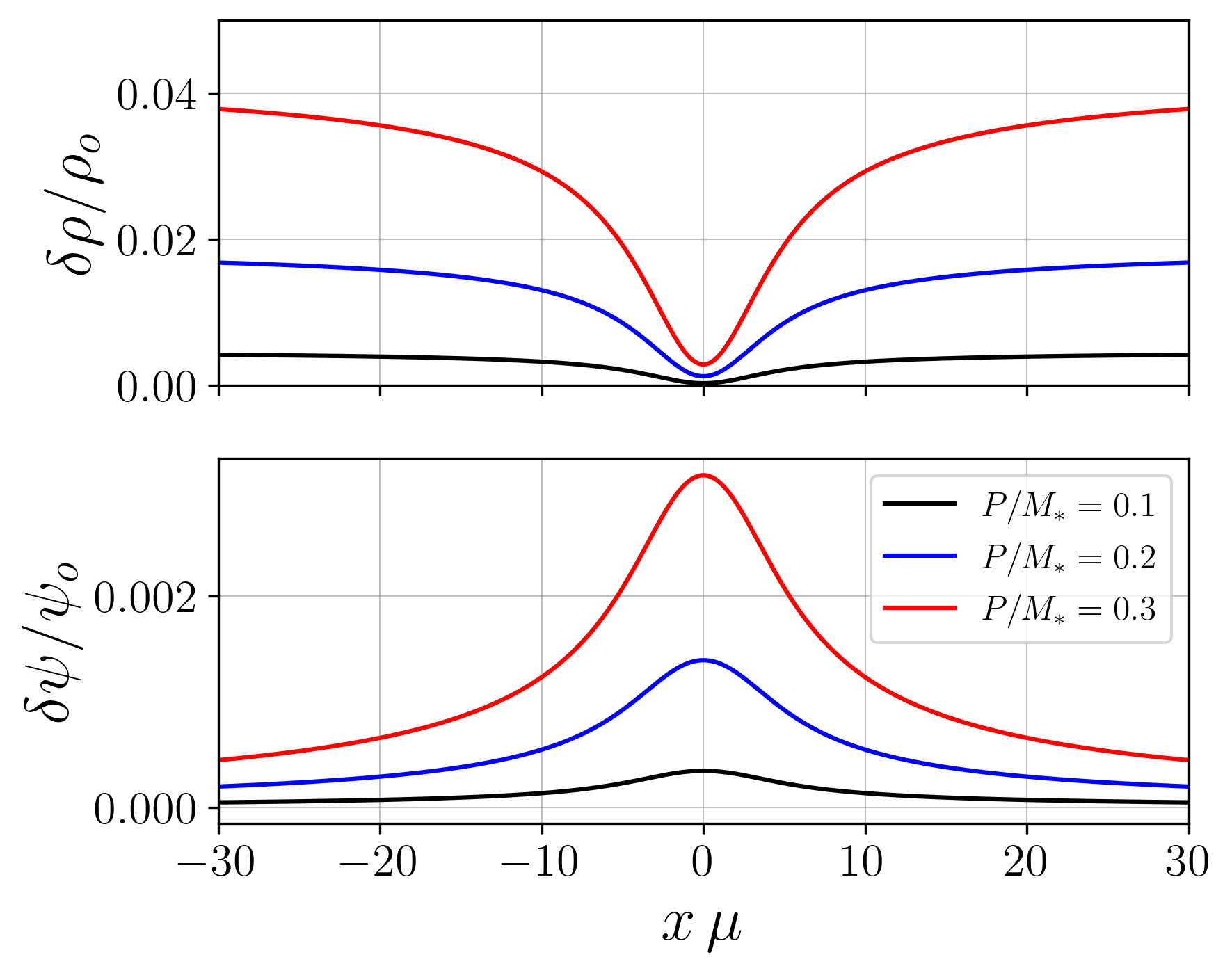} \\
\caption{Relative differences with respect to the stationary solution with vanishing momentum along the $x$ axis. The top panel shows the relative differences $\delta\rho/\rho_o$ in total mass-energy, and bottom panel shows those in the conformal factor $\delta\psi/\psi_o$. }
\label{fig:relativedifference}
\end{figure}

If we denote by $\varphi_o$, $\psi_o$, and $\alpha_o$ the stationary solutions of the amplitude of the scalar field, the conformal factor, and the lapse in isotropic coordinates, the initial data given by Eqs.~(\ref{eq:anzats}) and (\ref{eq:dot}) read
\begin{eqnarray}
\widetilde\Phi &=& \varphi_o\,e^{-i \frac{P\cdot x}{\mathcal{M}} }\\
\widetilde\Pi &=& - \frac{i\,\omega}{\psi_o^{-6}\alpha_o} \widetilde\Phi\,,\label{eq:initialPi}
\end{eqnarray}
where $\mathcal{M}$ is given by Eq.~(\ref{eq:calM}). Notice that we have chosen $\tilde\alpha = \psi_o^{-6}\alpha_o$ in order to recover the stationary solution when $P = 0$.
With these initial data, one can use Eqs.~(\ref{eq:KP2}) and (\ref{eq:Scalarrhoh2}) to construct the extrinsic curvature $\tilde{A}^{ij}$ and the source  $\rho_H$. Given these two, one solves the Hamiltonian constraint (\ref{eq:ham2}) for the conformal factor $\psi$ and obtains constraint-satisfying initial data.
Figure~\ref{fig:ADMmass} shows the initial ADM mass of the star as a function of $P/\Mstar$. One can show that for small momentum,  $M_{\text{ADM}} = \Mstar + \mathcal{O}(P^2)$, consistent with the growth observed in Fig.~\ref{fig:ADMmass}. To quantify the changes in the initial data that the momentum introduces, we plot in Fig.~\ref{fig:relativedifference} the relative differences with respect to the stationary solution with vanishing momentum. In the top panel is the relative difference for the total mass-energy density $\delta\rho/\rho_o$, and in the bottom panel is the one for the conformal factor $\delta\psi/\psi_o$ along the $x$ axis.

\begin{figure*}[!htbp]
\includegraphics[angle=0,width=0.95 \textwidth]{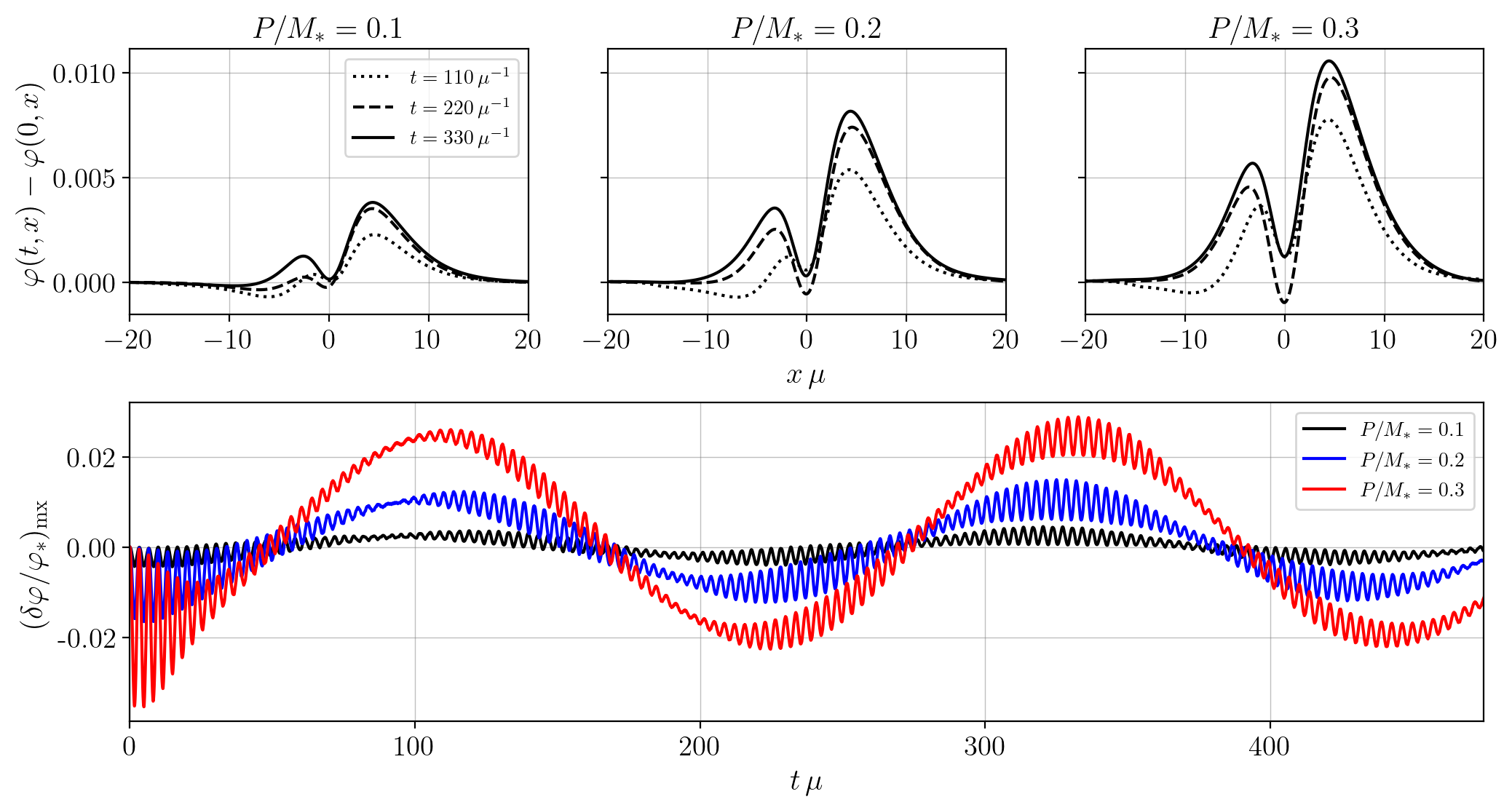}\\
\caption{For a single \bs{} with linear momentum, top panels depict, as a function of $x$, the change of the $\varphi$ profiles relative to the initial profile at times $t= 110,\, 220$ and $330\,\mu^{-1}$. Bottom panel displays the relative difference $(\delta\varphi/\varphi_*)_{\text{mx}}$ with respect to the initial maximum amplitude of the scalar field, $\varphi_*$, during the evolution.}
\label{fig:phiamp}
\end{figure*}

For the evolution, we use 6 levels of mesh refinement in a computational box of size $320\,\mu^{-1}$ with a grid spacing of $0.125\,\mu^{-1}$ for the finest mesh. As a sanity check, we evolved the case $P=0$ and obtained stationary solutions over several dynamical times. 

In our previous study of Bowen-York initial data evolutions with \ns{s}~\cite{Clark:2016ppe}, we observed that the \ns{} experiences oscillations in size and internal structure throughout its evolution. This is due to the assumption that the star is initially spherically symmetric, which does not account for deformations caused by the momentum. A similar situation occurs with our \bs{s}. In the top panels of Figure~\ref{fig:phiamp}, we show as a function of $x$ the change of the $\varphi$ profiles relative to the initial profile at times $t = 110,\, 220$ and $330\,\mu^{-1}$. Notice that the deformation of the profile is not symmetric; it is larger in amplitude and extent in the leading edge of the star along the direction of motion. This is counterintuitive since one would expect the opposite due to the Lorentz contraction. There is a different effect taking place since for the velocities we are considering, the Lorentz contraction is at most $\delta R_*/R_* < 0.05$. This effect is smaller than the deformation observed in the \bs{.}

We suspect that, as with the \ns{} case, the conformal flatness assumption is responsible for the observed perturbations in the star. Some evidence of that can be found in the oscillations in the maximum amplitude of $\varphi$. In the bottom panel of Fig.~\ref{fig:phiamp}, we present the relative difference $(\delta\varphi/\varphi_*)_{\text{mx}}$ with respect to the initial maximum amplitude of the scalar field, $\varphi_*$. There are clearly two oscillation modes, $\sigma_1$ and $\sigma_2$ that seem to enter as
\begin{eqnarray}
    (\delta\varphi/\varphi_*)_{\text{mx}} &=& A_1\,\sin(\sigma_1\,t+\theta_1)\nonumber\\
    &\times& [1+A_2\,\sin(\sigma_2\,t+\theta_2)]
\end{eqnarray}

\begin{figure}[h!]
\includegraphics[angle=0,width=0.45 \textwidth]{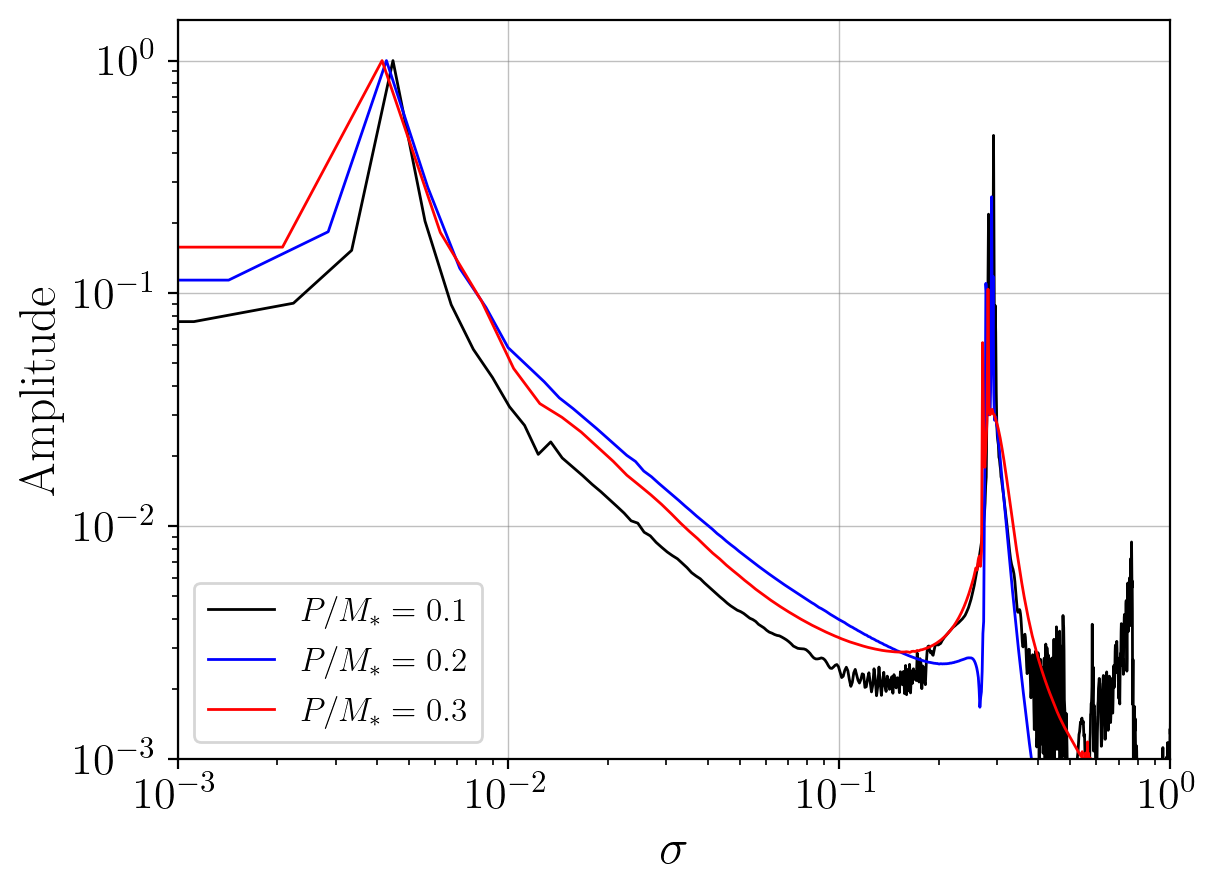} \\
\caption{Fourier transform of $(\delta\varphi/\varphi_*)_{\text{mx}}$ from the bottom panel of Fig.~\ref{fig:phiamp}.The two peaks correspond to frequencies $\sigma_1 = 0.00430\,\mu$ and $\sigma_2 = 0.2881\,\mu$.}
\label{fig:fft}
\end{figure}

Figure~\ref{fig:fft} shows the Fourier transform of $(\delta\varphi/\varphi_*)_{\text{mx}}$. The two modes are evident, $\sigma_1 \approx 0.00430\,\mu$ and $\sigma_2 \approx 0.2881\,\mu$. Since the modes do not exhibit a dependence on the momentum, it is safe to assume that these are normal modes of the \bs{.} Further evidence is found if we normalize the low-frequency mode by the average value of the lapse at the center of the star to account for the redshift. We get that $\sigma_1/\alpha = 0.0346\,\mu$, which is consistent with the value of $0.0374$ found in Ref.~\cite{HawleyChoptuik2000} for the same \bs{}.

\section{Head-on collisions of boson stars}
\label{sec:BBS}

The attractiveness of the Bowen-York method for compact object binary initial data with \ns{s} and \bs{s} is that it inherits the simplicity of the \bbh{} initial data. That is, the extrinsic curvature solutions to the momentum constraint are analytic and can be superposed since the momentum constraint is linear in the extrinsic curvature. The only added complication is in the source $\rho_H$ since it is not linear in the matter fields, see Eq.~(\ref{eq:Scalarrhoh2}). However, because the matter sources have compact support, or in the case of \bs{s}, approximately compact support, one can still carry out a superposition. 

Consider two \bs{s} at locations $x^i_+$ and $x^i_-$ with scalar amplitude $\varphi_{+}(x^i-x^i_+)$ and $\varphi_{-}(x^i-x^i_-)$ and linear momenta $P_{+}^i$ and $P_{-}^i$, respectively. We use the following superposition to construct $\rho_H$ and $\tilde A^{ij}$ for solving the Hamiltonian constraint:
\begin{eqnarray}
    \label{eq:\bbs{}_id}
    \tilde\alpha &=& \tilde\alpha_+ + \tilde\alpha_- - 1 \\
	\widetilde\Phi &=& \widetilde\Phi_+ + \widetilde\Phi_- \\
    \widetilde\Pi &=& \widetilde\Pi_+ + \widetilde\Pi_- \\
    \tilde{A}^{ij} &=& \tilde{A}_+^{ij} + \tilde{A}_-^{ij}\,.
\end{eqnarray} 

As with the coalescence of \ns{s}, \bs{} mergers could yield a \bs{} or a \bh{} depending on their masses. In Palenzuela et al.~\cite{Palenzuela:2006wp}, equal-mass head-on collisions of \bs{s} with vanishing initial momentum were studied. The initial data were constructed by superposing stationary solutions. The stars were initially sufficiently separated to minimize constraint violations. It was found that, for all cases in which the initial $\varphi_*>0.005$, a \bh{} forms upon collision. More recently~\cite{Helfer2022, Atteneder2024, Ge2025}, equal mass head-on collisions were investigated with $0.01\leq\varphi_*\leq0.08$ and boost velocities $v\leq0.1$. For all cases in which $\varphi_*\geq0.02$, the merger resulted in a \bh{}. The work in \cite{Ge2025} was part of a more comprehensive study to explore the two-dimensional parameter space of equal mass, non-spinning scalar solitonic \bs{s}. It has also been shown that for sufficiently high initial momenta, the stars behave as true solitons, passing through each other after colliding \cite{Lai2004}.

\begin{table}[!htbp]
\centering
\setlength{\tabcolsep}{12pt}
\renewcommand{\arraystretch}{1.3}
\begin{tabular}{ c c c c c c c c c}
\hline
\hline
 $\mathcal{C}$ & $\varphi_*$ & $\omega/\mu$ & $R_*\,\mu$ & $M_*\,\mu$ \\ 
\hline 
 0.028 & 0.02 & 0.954 & 17.25 & 0.475  \\
 0.039 & 0.03 & 0.933 & 13.78 & 0.542  \\
 0.050 & 0.04 & 0.913 & 11.67 & 0.584  \\
\hline
\hline
\end{tabular}
\caption{Parameters of the \bs{s} in isolation used in the simulations: Compactness $\mathcal{C}$, central amplitude $\varphi_*$, eigen-frequency $\omega$, areal radius $R_*$, and mass $M_*$.}
\label{tab:params}
\end{table}

We ran a series of nine \bbs{} head-on simulations. For each encounter, both stars have the same $\varphi_*$ and $P^i$. In Table~\ref{tab:params}, we list the parameters of the \bs{s} in isolation that we considered. We set initial momenta to $P/\Mstar = 0.1,\,0.2$, and $0.3$. All the simulations had an initial separation of $d = 80\,\mu^{-1}$. We use 6 levels of refinement, with a resolution $\Delta x = 0.125\,\mu^{-1}$ in the finest mesh. Waveforms were extracted at a radius $R = 330\,\mu^{-1}$. 

\begin{figure*}[!htbp]
\includegraphics[angle=0,width=0.95 \textwidth]{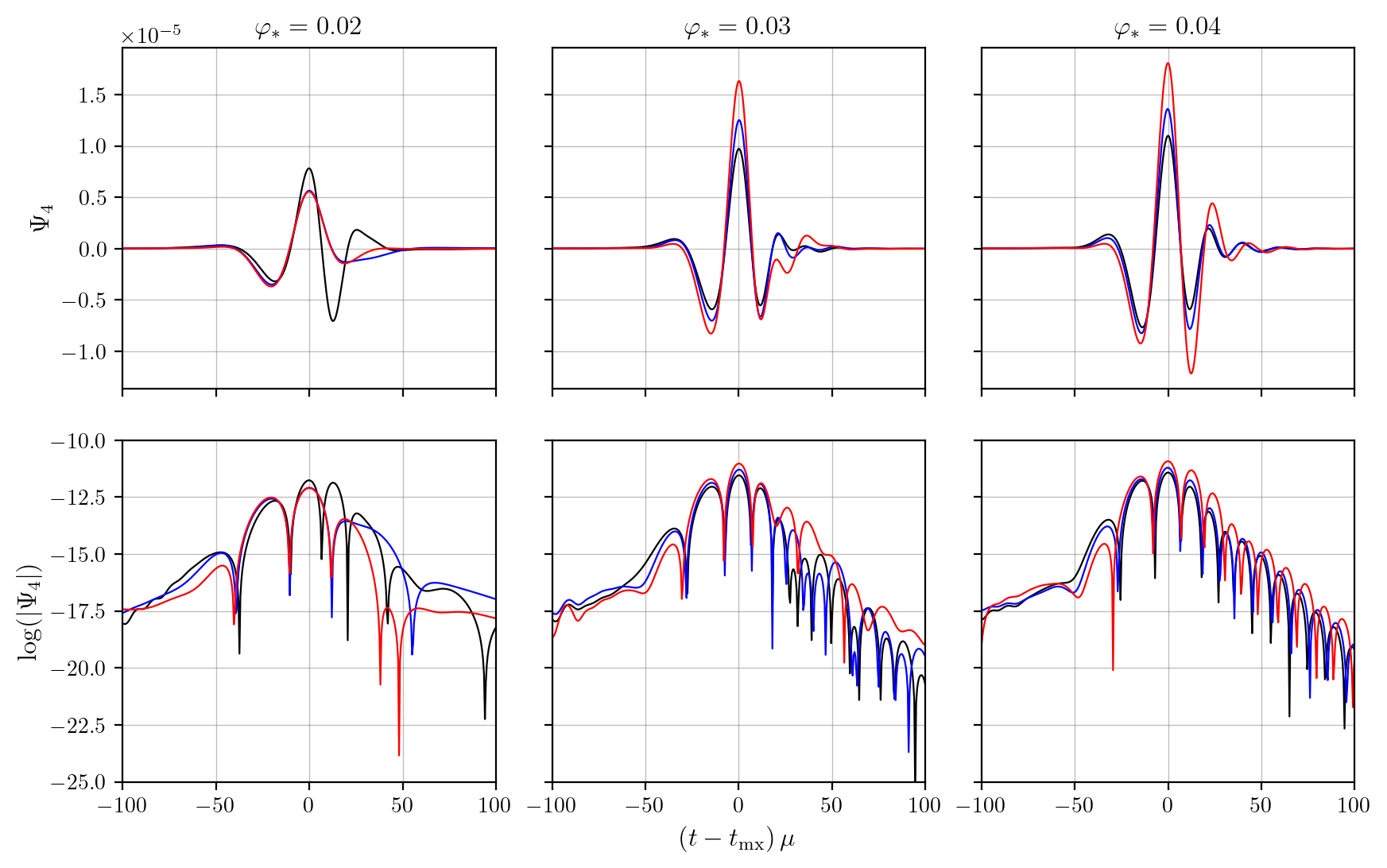}\\
\caption{Mode $l=2,\,m=0$ of the Weyl scalar $\Psi_4$ for all the simulations, with lower panels showing $|\Psi_4|$.  Panels in rows from left to right are for $\varphi_*= 0.02,\,0.03$ and 0.04, respectively. Black, blue, and red lines correspond to $P/\Mstar = 0.1,\,0.2$, and 0.3, respectively.}
\label{fig:psi4BBS}
\end{figure*}

In Figure~\ref{fig:psi4BBS}, we show the $l=2,\,m=0$ mode of the Weyl scalar $\Psi_4$ for all the simulations. To facilitate the identification of \qnm{} ringing, the bottom panels show $|\Psi_4|$. The outcome of the collisions with $\varphi_* = 0.03$ and 0.04 was a \bh{} for all the momenta values considered. These \bs{s} have high compactness (see Table~\ref{tab:params}). Presence of a \bh{} is evident from the characteristic \qnm{} sinusoidal exponential damping observed in the bottom panels. We also got unquestionable evidence for the formation of a \bh{} from the detection of an apparent horizon after the merger.

\begin{figure}[h!]
\includegraphics[angle=0,width=0.4 \textwidth]{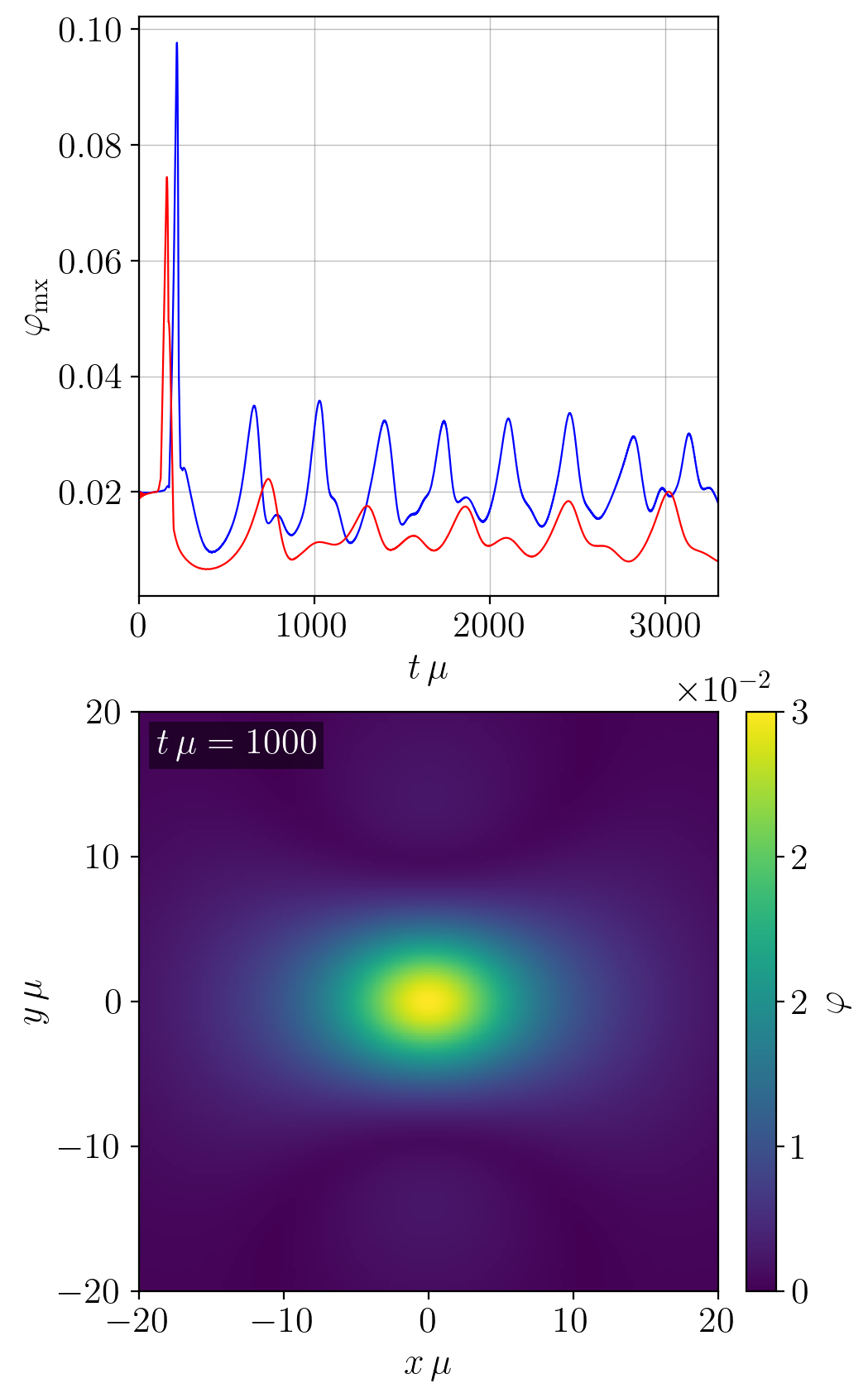} \\
\caption{Top panel depicts the maximum of the scalar field amplitude $\varphi_{\text{mx}}$ for the case $\varphi_*=0.02$ and $P/M_*=0.2$ (blue), and $P/M_*=0.3$ (red). The bottom panel show a snapshot of $\varphi$ in the $x$-$y$ plane at time $t=1000\,\mu^{-1}$ for $P/M_*=0.2$.}
\label{fig:BBS2}
\end{figure}

The case $\varphi_*=0.02$ is peculiar. For sufficiently low momentum, one expects \bh{} formation, in agreement with Palenzuela et al.~\cite{Palenzuela:2006wp}. This was indeed the case for $P/\Mstar = 0.1$. However, for momenta $P/\Mstar=0.2$ and 0.3, the outcome was a highly distorted and oscillating \bs{}. A combination of high kinetic energy and low compactness produced a splash at merger large enough to prevent collapse into a \bh{}, but not strong enough to completely disrupt the star. A better understanding of the situation can be found in Figure~\ref{fig:BBS2}. In the top panel, we show the maximum value of the scalar field amplitude $\varphi_{\text{mx}}$ for the case $\varphi_*=0.02$ with $P/M_*=0.2$ and 0.3. The spike occurs when the stars merge. After that, one has the oscillations of the highly distorted remnant \bs{}. Comparing the case $P/M_*=0.3$ with $P/M_*=0.2$, we observe that the former has a smaller spike at merger, and the amplitude of oscillations after merger is smaller. From the average value of $\varphi_{\text{mx}}$ after merger, we have that $\varphi_* \approx 0.0207$ for $P/M_*=0.2$ and $\varphi_* \approx 0.0122$ for $P/M_*=0.3$. The fundamental oscillation frequencies for \bs{s} with these central values are $0.0028\,\mu^{-1}$ and $0.0017\,\mu^{-1}$, approximately the same as the colliding stars, and consistent with the frequencies observed in the middle panel of Fig.~\ref{fig:BBS2}. The bottom panel in Fig.~\ref{fig:BBS2} shows a snapshot of $\varphi$ at a time $t = 1000\,\mu^{-1}$.

\begin{table}[!htbp]
\centering
\renewcommand{\arraystretch}{1.2}
\setlength{\tabcolsep}{3.3pt}
\begin{tabular}{ c | c | c | r | c | c | c}
\hline
\hline
        $\varphi_*$ & $P/M_*$ & $M_{\text{adm}}\,\mu$ &  $E\,\mu\,10^{-4}\,(\%)$  &  $\Mhole\,\mu$  &  $\omega_q\,\Mhole$ &  $\tau_q/\Mhole$       \\
\hline
       0.02 & 0.1 & 0.952 &3.21 (0.03) &    0.93   &     0.06 & 28.03     \\ 
       0.02 & 0.2 & 0.966 &4.33 (0.04)&    -       &     -     & -          \\ 
       0.02 & 0.3 & 0.990 &5.29 (0.05)&    -       &     -     & -          \\ 
       0.03 & 0.1 & 1.087 &3.64 (0.03)&    1.08   &     0.45 & 11.60     \\ 
       0.03 & 0.2 & 1.103 &6.26 (0.06)&    1.07   &     0.44 & 10.96     \\ 
       0.03 & 0.3 & 1.129 &13.03 (0.11)&    1.04   &     0.21 & 10.57     \\
       0.04 & 0.1 & 1.170 &4.76 (0.04)&    1.16   &     0.36 & 11.26     \\ 
       0.04 & 0.2 & 1.188 &7.22 (0.06)&    1.17   &     0.37 & 11.36     \\ 
       0.04 & 0.3 & 1.216 &14.81 (0.12)&    1.19   &     0.37 & 10.49     \\
       
\hline
\hline
\end{tabular}
\caption{ADM energy $M_{\text{adm}}$,  energy emitted $E$ and its \% of ADM energy, remnant \bh{} mass $\Mhole$, \qnm{} frequency $\omega_q$ and decay time scale $\tau_q$.}
\label{tab:finalBH}
\end{table} 

The mass of the final \bh{} was obtained from the apparent horizon. Fits to $\Psi_4\propto \exp{(-t/\tau_q)}\,\sin{(\omega_q\,t+\phi)}$ gave the \qnm{} frequency $\omega_q$ and decaying time scale $\tau_q$. 
In Table~\ref{tab:finalBH}, we report these quantities. Since the head-on involves non-rotating \bs{s}, the final \bh{}, if formed, is non-spinning, i.e., a Schwarzschild \bh{}. The \qnm{} frequency and decay time for a Schwarzschild \bh{} are $\omega_q\,\Mhole=0.3737$ and $\tau_q/\Mhole=11.24$. Looking at the results in Table~\ref{tab:finalBH} for $\varphi_* = 0.03$ and 0.04, we see that the values of $\tau_q$ are relatively close to those of a Schwarzschild \bh{}. For the \qnm{} frequency, some cases are also close to Schwarzschild values, but others differ significantly, e.g. $\varphi_* = 0.03$ with $P/\Mstar = 0.3$. The reason for these differences is that in the vicinity of the \bh{} there is a remnant scalar field that is being accreted. This situation is much more extreme in the case $\varphi_* = 0.02$ with $P/\Mstar = 0.1$ where the fitting gave significantly different values from the Schwarzschild \bh{}. By looking at the waveform for this case in Fig.~\ref{fig:psi4BBS} (black line), the absence of a clean \qnm{} ringing is evident. 
A similar persistent quasi-bound state formed around a \bh{} that contaminates \qnm{} ringing has been observed in the spherically symmetric collapse of unstable \bs{s}~\cite{Esco2017}. 

Finally, as expected, the energy radiated in gravitational waves increases with both the central amplitude and the initial momenta, in agreement with the results from \cite{Ge2025}. 

\section{Head-on collisions of boson stars with black holes}
\label{sec:BHBS}

Head-on collisions of \bs{s} with \bh{s} have received far less attention. In \cite{Cardoso2022}, a \bs{} of central amplitude $\varphi_*=0.02$ was fixed at the origin while a \bh{} of varying mass and initial velocity pierced through it. Even for a highly boosted \bh{} with a mass significantly smaller than that of the BS, the \bh{} accretes almost the entire BS. Recently~\cite{Marks2026}, head-on collisions of \bhbs{} systems were investigated across the entire stable branch ($\varphi_* \leq 0.076$), with encountered velocities as high as $v=0.7$ for equal mass systems and velocities $v=0.1$ for unequal. The study found that the \bhbs{} encounters always radiate less \gw{} energy than their \bbh{} counterparts. 

 Our \bhbs{} head-on collisions are equal mass, with a non-spinning \bh{}. To make a reasonable comparison with the \bbs{} collisions of the previous section, the same parameters listed in Table~\ref{tab:params} for the case $\varphi_*=0.04$ were used. As before, we ran a series of nine head-on simulations, three each for \bhbs{}, \bbs{}, and \bbh{}. For each binary type, we set initial momenta to $P/\Mstar = 0.1,\,0.2$, and $0.3$. All the simulations had an initial separation of $d = 80\,\mu^{-1}$. We use 8 levels of refinement for the collisions involving \bh{s}, with a resolution $\Delta x = 0.03125\,\mu^{-1}$ in the finest mesh. Waveforms were extracted at a radius $R = 330\,\mu^{-1}$. For the \bhbs{} head-on collisions, the source function $\rho_H$ requires only $\widetilde\Phi$ and $\widetilde\Pi$ from the \bs{}. On the other hand, the traceless extrinsic curvature has two contributions:
$\tilde{A}^{ij} = \tilde{A}_{\text{bs}}^{ij} + \tilde{A}_\text{h}^{ij}$, with $\tilde A^{ij}_{\text{h}}$ constructed from Eq.~(\ref{eq:KP}) and $\tilde A^{ij}_{\text{bs}}$ from Eq.~(\ref{eq:KP2}).

\begin{figure*}[!htbp]
\includegraphics[angle=0,width=0.98 \textwidth]{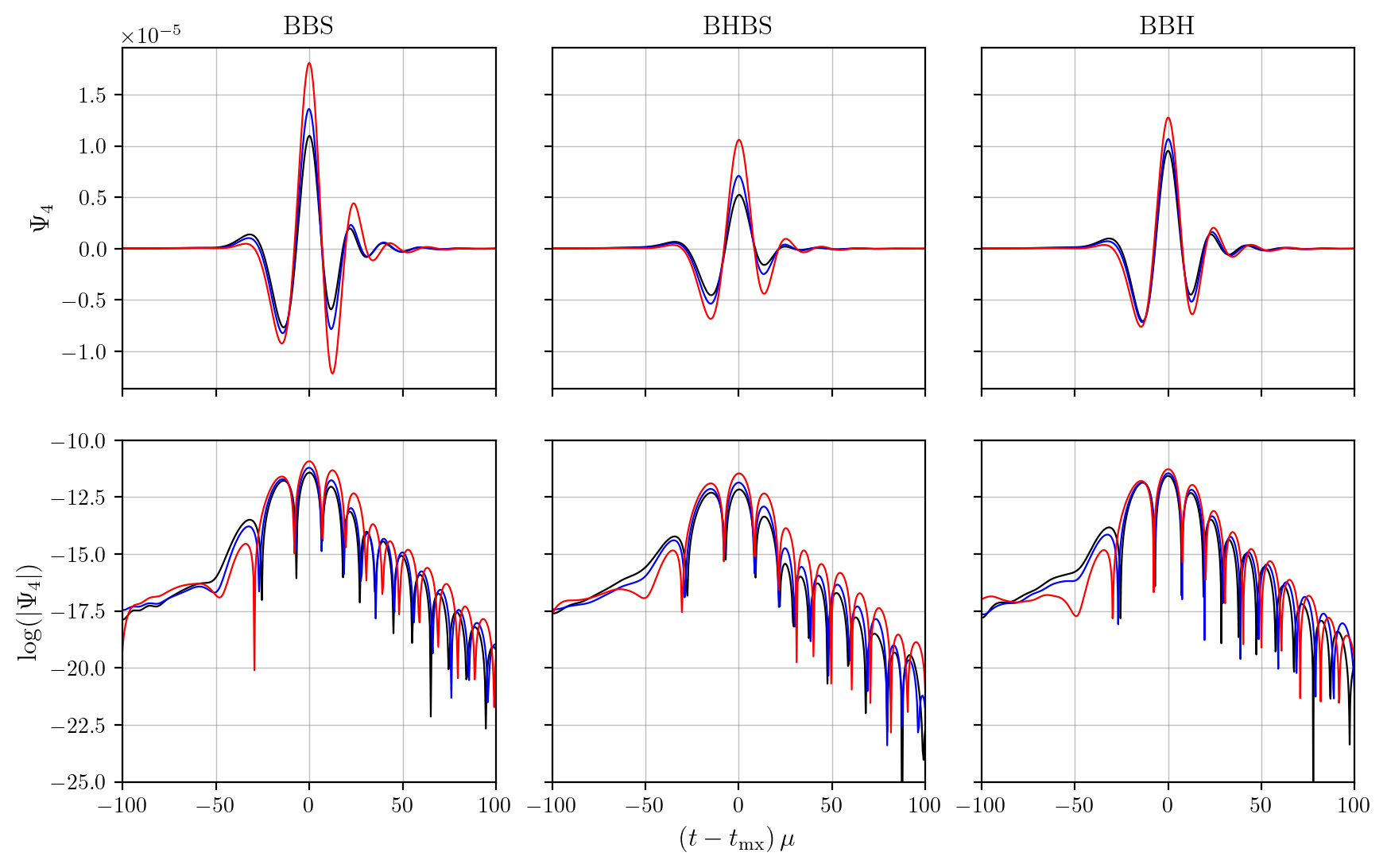}\\
\caption{Mode $l=2,\,m=0$ of the Weyl scalar $\Psi_4$, with lower panels showing $|\Psi_4|$.  Panels in rows from left to right are for \bbs{}, \bhbs{}, and \bbh{}, respectively. Black, blue, and red lines correspond to $P/\Mstar = 0.1,\,0.2$, and 0.3, respectively.}
\label{fig:psi4BHBS}
\end{figure*}

Figure~\ref{fig:psi4BHBS} shows the mode $l=2,\,m=0$ of the Weyl scalar $\Psi_4$, with lower panels showing $|\Psi_4|$. Panels in rows from left to right are for \bbs{}, \bhbs{}, and \bbh{}, respectively. Black, blue, and red lines correspond to $P/\Mstar = 0.1,\,0.2$, and 0.3, respectively. During the pre-merger phase, the waveforms show similar shape characteristics. The differences are in the magnitude of the amplitudes. From largest to smallest amplitudes are \bbs{}, \bbh{}, and \bhbs{}. This is naturally reflected in the energy emitted in \gw{s}, as seen in Table~\ref{tab:BHBS}. For a given linear momentum, the \bbs{} radiate the largest percentage of the initial total ADM energy, while the \bhbs{} radiate the least. It is quite surprising that the \bbs{} mergers emit more energy than the equivalent \bbh{}. This was also found in \cite{Ge2025}, in which it was posited that the collision of the less compact \bs{s} produces a more perturbed \bh{} remnant. This could also explain why the \bhbs{} collisions produce the least radiated energy, as the deformation of the \bh{} is not as significant compared to the other cases. Our results are also in agreement with those of \cite{Marks2026}, where it was found that \bhbs{} head-on collisions are always less efficient emitters than their \bbh{} counterparts.

From the bottom panels of Fig.~\ref{fig:psi4BHBS}, it is evident that the outcome of all collisions is a ringing \bh{} since the waveform has the ubiquitous sinusoidal exponential damping. 
In Table~\ref{tab:BHBS}, we show the \qnm{} frequency $\omega_q$ and decaying timescale $\tau_q$ from fittings of $\Psi_4 \propto \exp(-t/\tau_q)\sin{(\omega_q t + \phi)}$, and all values obtained are in close agreement with the respective \qnm{} frequency and decay time for a Schwarzschild \bh{}. Interestingly, the decay timescale for \bbs{} is closer to the true value than that for \bhbs{}.

\begin{table}[!htbp]
\centering
\renewcommand{\arraystretch}{1.2}
\setlength{\tabcolsep}{2.7 pt}
\begin{tabular}{ l | c | c | r | c | c | c }
\hline
\hline
        Type & $P/M_*$ & $M_{\text{adm}}\,\mu$ &    $E\,\mu\,10^{-4}\,(\%)$ &$\Mhole\,\mu$  &  $\omega_q\,\Mhole$ &  $\tau_q/\Mhole$       \\
\hline
       \bbs{}  &0.1 & 1.170 & 4.76 (0.04) & 1.16 & 0.36 & 11.26 \\ 
       \bbs{}  &0.2 & 1.188 & 7.22 (0.06)& 1.18 & 0.37 & 11.36 \\ 
       \bbs{}  &0.3 & 1.216 & 14.81 (0.12)& 1.19 & 0.37 & 10.49 \\
       \bhbs{} &0.1 & 1.179 & 2.01 (0.02)& 1.17 & 0.36 & 12.87 \\ 
       \bhbs{} &0.2 & 1.198 & 3.36 (0.03)& 1.19 & 0.37 & 12.89 \\ 
       \bhbs{} &0.3 & 1.230 & 8.13 (0.07)& 1.22 & 0.38 & 12.95 \\ 
       \bbh{}  &0.1 & 1.182 & 3.92 (0.03)& 1.17 & 0.37 & 10.68 \\ 
       \bbh{}  &0.2 & 1.203 & 5.21 (0.04)& 1.19 & 0.37 & 11.37 \\
       \bbh{}  &0.3 & 1.237 & 9.56 (0.08)& 1.23 & 0.38 & 11.98 \\
\hline
\hline
\end{tabular}
\caption{ADM energy $M_{\text{adm}}$,  energy emitted $E$ and its \% of ADM energy, remnant \bh{} mass $\Mhole$, \qnm{} frequency $\omega_q$ and decay time scale $\tau_q$.}
\label{tab:BHBS}
\end{table}

\section{Conclusions}
\label{sec:conclusions}
The present study investigated the effectiveness of the Bowen-York type initial data method applied to binary systems with \bs{} components. We first tested the method on a single-\bs{} with linear momentum and saw that the assumption of conformal flatness and spherical symmetry in the star triggers oscillations in the \bs{}, similar to those observed in our study of Bowen-York initial data for \ns{s}~\cite{Clark:2016ppe}. As a first step, we focused on head-on collisions for a range of encounter velocities. In general terms, the results of the energy emitted, and the ringdown analysis of \gw{s} are in agreement with previous studies. In particular, we corroborated that \bbs{} systems radiate more than the equivalent \bbh{} systems, but \bhbs{} systems radiate less energy than their \bbh{} counterparts. We also found that for large encounter velocities, \bbs{} systems leave behind an oscillating \bs{}. The larger the initial momentum, the more the final \bs{} differs from its progenitors.

In a subsequent study,  we will apply this initial data prescription to inspiraling compact object binary coalescences with \bs{s}, including mergers with spinning \bh{s} and \ns{s}. As with \bbh{s}, the Bowen-York recipe will allow us to make a connection with the early, post-Newtonian inspiral phase in order to set up astrophysically relevant initial data for numerical relativity simulations.

\paragraph*{Acknowledgements}
This work is supported by NSF grants PHY-2114582 and PHY-2207780. 

\vspace{0.25in}
\bibliographystyle{iopart-num}

\providecommand{\newblock}{}

\end{document}